# Renormalized Effective Quantum Number for Centrally Symmetric Problems

N.N.Trunov[1]

*D.I.Mendeleyev Institute for Metrology*

*Russia, St.Petersburg. 190005 Moskovsky pr. 19*

(Dated: February 25, 2010)

The universal effective quantum number that determines the level ordering in arbitrary centrally symmetric potentials is defined more precisely by means of an improved variant of the semiclassical approach

## 1. Introduction

We have earlier introduced and calculated a new universal effective quantum number which determines the level ordering and other properties of the centrally symmetric systems, though it is not formally an exact quantum number[1]. In the most cases a linear combination of the radial and orbital quantum numbers is exact enough.

On the one hand, we have used the usual semiclassical approximation in a certain combination wih other methods. On the other hand, we have developed a new semiclassical approach with two parameters [2-4]. In what follows we combine both approaches in order to refine the effective quantum number.

## 2. Effective quantum number

In this section we remind some results of [1].

We start from the Schrödinger equation for a centrally symmetric potential $U(r) < 0$

---

[1] Electronic address: trunov@vniim.ru

$$\hbar \Delta \Psi + 2(E - U(r)) = 0 \tag{1}$$

assuming usual for atomic problems conditions

$$r^2 U(r) \to 0, \text{ if } r \to 0, \infty \tag{2}$$

Hereafter we only study the appearance of new bound states when its energy $E = 0$, though the whole approach is valid for any $E$.

For solving (1) we intend to use a semiclassical approach. As it was shown in [1], best results may be achieved in a new conformal metric which corresponds to a new variable $\rho = \ln r$. Taking into account the curvature of this metric, we obtain for $E = 0$ the radial equation ($m = 1$)

$$\hbar^2 \Psi'' + [W(\rho) - \lambda^2] \Psi = 0, \tag{3}$$

$$W(\rho) = -2e^\rho U(e^\rho) \geq 0, \tag{4}$$

$$\lambda = l + \frac{d-2}{2} \tag{5}$$

with $l$ being the orbital quantum number and $d$ the dimensionality of the space; usually $d = 3$, but for new physical nanosystems may be another $d$.

From (2) follow limiting values

$$W(-\infty) = W(\infty) = 0 \tag{6}$$

Usual semiclassical approximation leads to

$$I(\lambda) = \frac{1}{\pi \hbar} \int \sqrt{W - \lambda^2} \, d\rho = n + \tfrac{1}{2} \equiv \nu \tag{7}$$

where $n$ is the radial quantum number. The integration limits in (7) and in what follows are the corresponding turning points.

Our condition (7) is now inconvenient since it contains $\lambda$ inside the integral. Introducing an unknown yet function $t(\lambda)$ such that

$$t(0) = 0, \ t(\lambda) > 0 \text{ if } \lambda > 0 \tag{8}$$

we always can write:

$$I(\lambda) = I(0) - t(\lambda), \tag{9}$$

then instead of (7) we obtain

$$\frac{1}{\pi\hbar}\int\sqrt{W}d\rho = T(\nu,\lambda) \qquad (10)$$

$$T(\nu,\lambda) = \nu + t(\lambda) \qquad (11)$$

Here $T(\nu,\lambda)$ is our new universal effective quantum number for centrally symmetric problems. As it was discussed and calculated in [1], a simple linear approximation for $t(\lambda)$, i.e.

$$T(\nu,\lambda) = \nu + \varphi\lambda \qquad (12)$$

has a quite sufficient accuracy for the most situations. Several ways for calculating $\varphi$ lead to very close values; they depend certainly on a given $U(r)$. Values for a number of potentials, universal diagrams determining the level ordering and other applications may be found in [1] (including delicate results which need a nonlinear refined $t(\lambda)$).

Note that if we introduce a new function – "potential" $V$ and a new variable – "energy" $\varepsilon$ such that

$$2V(\rho) = V_m - W(\rho) \geq 0, \qquad (13)$$

$$V_m = \max_\rho W(\rho), \qquad (14)$$

$$2\varepsilon = V_m - \lambda^2 \geq 0, \qquad (15)$$

our equation (3) takes the usual form

$$\hbar\Psi'' + 2[\varepsilon - V]\Psi = 0 \qquad (16)$$

From (6) follow limiting conditions

$$V_m = V(-\infty) = V(\infty) \qquad (17)$$

$$0 \leq \varepsilon \leq V_m \qquad (18)$$

We must certainly keep in mind that $\varepsilon$ is only a formal parameter (the genuine $E = 0$ in the present paper). In intermediate calculations we can treat $\lambda$ as a continuous variable.

## 3. Improved semiclassical approach

We can always write an equation for the exact spectrum of (16), (17) as

$$\Phi(\varepsilon) = \frac{\sqrt{2}}{\pi\hbar}\int\sqrt{\varepsilon_n - V(x)}\,dx = n + \tfrac{1}{2} + \delta(\varepsilon) \qquad (19)$$

with $n = 0,1,..$ and an unknown yet function $\delta(\varepsilon)$ ensuring exact values of $\varepsilon_n$.

It is known a series for $\delta$:

$$\delta(\varepsilon) = \sum_{k=1}^{\infty} \delta_k(\varepsilon) \qquad (20)$$

where the $k$-th term is proportional to $\hbar^{2k-1}$,

$$\delta_1 = \frac{\hbar}{24\pi}\frac{\partial^2}{\partial\varepsilon^2}\int\frac{dx}{\sqrt{\varepsilon-V}}\left(\frac{dV}{dx}\right)^2 \qquad (21)$$

Another $\delta_k$ look as cumbersome integro-differential operators acting on $V$ and are not practically used. In the most cases one also put $\delta_1 = \delta = 0$.

We have noticed earlier [2] that a very wide class of potentials can be expressed by means of an auxiliary function $s(x)$ so that

$$\begin{aligned} V(x) &= P_2(s), \\ \frac{ds}{dx} &= Q_2(s) \end{aligned} \qquad (22)$$

with quadratic polynomials $P_2$, $Q_2$, the series (20) can be easily summarized [3,4]. Namely, for (22) $\delta_n$ do not depend on $\varepsilon$ and

$$\delta = \frac{2\delta_1}{1+\sqrt{1+16\delta_1^2}} = \frac{\sqrt{1+16\delta_1^2}-1}{8\delta_1}. \qquad (23)$$

Limits of (23) as a function of $\delta_1$ are

$$\delta = \delta_1 \quad \text{if } |\delta_1| \ll 1 \qquad (24)$$

$$\delta = \frac{\operatorname{sgn}\delta_1}{2} - \frac{1}{8\delta_1}, \quad |\delta_1| \gg 1 \qquad (25)$$

It is known that in all potentials with coinciding asymptotics at $\pm\infty$ like (17) the lowest bound state with $n = 0$ exists for each value of its depth $V_m$. If $V_m \to 0$, $\varepsilon_0$ practically coincides with $V_m$.

Putting $n = 0$ in (19) we see that the left and the right sides are equal only if

$$\delta = -\frac{1}{2} + \Phi(V_m) \tag{26}$$

Comparing (25) and (26) we see that:

$$\delta_1 = -\frac{1}{8\Phi_m}; \quad \Phi_m = \Phi(V_m) \tag{27}$$

Substituting $\delta$ (23) with $\delta_1$ (27) into (19) we obtain

$$\Phi(\varepsilon) = n + \frac{1}{2} + \Phi_m \left[ 1 - \sqrt{1 + \frac{1}{4\Phi_m^2}} \right] \tag{28}$$

If we trace when a new bound state with $\varepsilon_n = V_m$ appears, we have to put $\Phi(\varepsilon) = \Phi(V_m) = \Phi_m$ in the left side of (28). The final expression may be rewritten in a simple form equivalent to (28):

$$\Phi_m = \left[ v^2 - \frac{1}{4} \right]^{1/2}; \quad v = n + \frac{1}{2} \tag{29}$$

We see that the lower bound state with $v = \frac{1}{2}$ really exists even if $\Phi_m \to 0$.

In a general case of potentials (20) and (21) are some smooth functions of our "energy", but all the above results maintain a high accuracy since 1) both limits (25) remain valid and 2) genuine physical potentials are close in some adequate metric to (22).

## 4. Renormalization of the effective quantum number

Now we return to our radial problem (section 2). Comparing (7) and (19) expressed by means of formal parameters $\varepsilon$, $V$ we obtain:

$$I(\lambda) = \Phi(\varepsilon) \tag{30}$$

$$\Phi_m = \Phi(V_m) = I(0) \tag{31}$$

Taking into account (9), we have:

$$\Phi(\varepsilon) = \Phi_m - t(\lambda) \tag{32}$$

or instead of (28)

$$\Phi_m = T(\nu,\lambda) + \Phi_m\left[1 - \sqrt{1 + \frac{1}{4\Phi_m^2}}\right] \tag{33}$$

The final form we obtain by analogy to (29)

$$\frac{1}{\pi\hbar}\int \sqrt{W}\,dp = T_{ren}(\nu,\lambda) \tag{34}$$

$$T_{ren}(\nu,\lambda) = \left[T^2(\nu,\lambda) - \tfrac{1}{4}\right]^{1/2} \tag{35}$$

with $T$ (11).

Our renormalized effective quantum number $T_{ren}$ determines appearing of new levels $(n,l)$ with growing $V_m$ more precise than $T$.

Replacement of $T$ by $T_{ren}$ does not change the level ordering:

$$T_{ren}(\nu,\lambda) > T_{ren}(\nu',\lambda') \leftrightarrow T(\nu,\lambda) > T(\nu',\lambda') \tag{36}$$

but the corresponding value of $V_m$ decreases, especially for lower states with small $\nu, \lambda$.

For a linear $t(\lambda)$ we obtain:

$$T_{ren} = \left[(\nu + \varphi\lambda)^2 - \tfrac{1}{4}\right]^{1/2} \tag{37}$$

Two first terms of the power series expansion (37) are:

$$T_{ren} = \nu + \varphi\lambda - \frac{1}{8(\nu + \varphi\lambda)} = T - \frac{1}{8T} \tag{38}$$

so that an addition to $T$ decreases rather rapidly by increasing $T$.

## 5. Conclusion

Thus, we have constructed an improved renormalized effective quantum number (35), (37) for centrally symmetric problems. Though the previous $T$ (11) is exact enough in the most cases, new corrections are useful since we can evaluate the accuracy of $T$ and to correct appearing of the lower bound states. It may be important e.g. for the structure of the Periodic system of elements (see [1] and references there).

It is known a family of so-called Lenz potentials

$$V_a(r) = -\frac{Z}{r^2 \left(r^a + r^{-a}\right)^2} \tag{39}$$

satisfying conditions (2) if $a > 0$. A new level with $E = 0$ appears by increasing $v$ when

$$Z = 2a\left[\left(v + \frac{\lambda}{a}\right)^2 - \frac{1}{4}\right]^{1/2} \tag{40}$$

The same exact result we obtain calculating $\Phi_m$ for (39).

For example, we can minimize the mean-square deviation

$$\frac{d}{d\varphi}\int d\lambda \left[I(\lambda) - I(0) + \varphi\lambda\right]^2 = 0 \tag{41}$$

in order to determine $\varphi$; more elaborate approaches [1] lead to the same value $\varphi = 1/a$.

For $a = 1/2$ we obtain the Tietz potential

$$V = -\frac{Z}{r(1+r)^2} \tag{42}$$

which was earlier used as a simple approximation of the Thomas-Fermi (TF) potentials. However we have noticed that its $\varphi = 2$ does not coincide with the real for the Periodic system as well as for an individual atom (and for the TF ) value $\varphi \approx 7/4$.

In a general case various methods lead to not coinciding but very close results so that accuracy remains quite sufficient.